\documentclass[twocolumn,showpacs,superscriptaddress,prl]{revtex4}

\usepackage{amsmath, amsthm, amssymb}
\usepackage{graphicx}
\usepackage{dcolumn}
\usepackage{bm}
\usepackage{color} 
\usepackage{ulem}

\newcommand{\ket}[1]{| #1 \rangle}

\newcommand{\tr}{\mathrm{Tr}}

\begin{document}

\title{Gapped Two-Body Hamiltonian for continuous-variable quantum computation}

\author{Leandro Aolita}
\affiliation{ICFO-Institut de Ci\`{e}ncies Fot\`{o}niques, Mediterranean
Technology Park, 08860 Castelldefels (Barcelona), Spain}

\author{Augusto J. Roncaglia}
\affiliation{ICFO-Institut de Ci\`{e}ncies Fot\`{o}niques, Mediterranean
Technology Park, 08860 Castelldefels (Barcelona), Spain}

\author{Alessandro Ferraro}
\affiliation{Grup d'\`{O}ptica, Universitat Aut\`{o}noma de Barcelona, E-08193 Bellaterra (Barcelona), Spain}

\author{Antonio Ac\'in}
\affiliation{ICFO-Institut de Ci\`{e}ncies Fot\`{o}niques, Mediterranean
Technology Park, 08860 Castelldefels (Barcelona), Spain}
\affiliation{ICREA-Instituci\'o Catalana de Recerca i Estudis
Avan\c cats, Lluis Companys 23, 08010 Barcelona, Spain}

\begin{abstract}
We introduce a family of Hamiltonian systems for measurement-based quantum computation with continuous variables. 
The Hamiltonians ({\it i}) are quadratic, and therefore two body, ({\it ii}) are of short range, ({\it iii}) are frustration-free, and ({\it iv}) possess  a constant energy gap proportional to the squared inverse of the squeezing. Their ground states are the celebrated Gaussian graph states,  which are universal resources for quantum computation in the limit of infinite squeezing.  These Hamiltonians constitute the basic ingredient for the adiabatic preparation of graph states and thus open new venues for the physical realization of continuous-variable quantum computing beyond the standard optical approaches. We characterize the correlations in these systems at thermal equilibrium. In particular, we prove that the correlations across  any multipartition are contained exactly in its boundary, automatically yielding a correlation area law.  
\end{abstract}

\pacs{03.67.Ac,03.67.Lx,42.50.-p}

\maketitle
{\it Introduction.---}The realization of  a device that can perform arbitrary quantum-state manipulations -- a universal quantum processor -- is nowadays one of the most active and promising searches in physics.  All the resources on which we count for the realization of these machines in turn fall into one out of two fully-general categories: quantum states of discrete-variable (DV), finite-dimensional systems and quantum states of continuous-variable (CV), infinite-dimensional ones. For both categories, considerable  progress  has been achieved. In particular, an entire spectrum of new possibilities for state manipulation was opened by the landmark discovery that it is possible to process quantum information in a universal manner by the simple act of performing local measurements  \cite{Raussendorf01}. This discovery originally took place in the finite-dimensional scenario, but was later on extended to CV systems \cite{Menicucci}. In these measurement-based quantum-computation (MBQC) approaches, information processing proceeds by a sequence of adaptive single-particle measurements on  massively entangled multiparticle states prepared in advance.  These states are the so-called cluster states, introduced first for DV systems \cite{Raussendorf01,clusterD} and then extended to the CV case \cite{Menicucci,Zhang}. The local measurements consume cluster-state entanglement as the main resource of the computation. 

\par Cluster states are in turn particular instances  of a more general family: the graph states \cite{graph_review}. Other  examples of   graph states of importance are the Greenberger-Horne-Zeilinger states and many codewords for quantum error-correction \cite{Raussendorf07}. Every graph state is  associated to a mathematical graph $G_{(\mathcal{V},\mathcal{E})}\equiv\{\mathcal{V},\mathcal{E}\}$, of vertices $i\in\mathcal{V}$ and edges  $\{i,j\}\in\mathcal{E}$, with $1\leq i,\ j\leq N$. In both the DV and CV cases, graph states can be created in an operational way: starting from a product state of $N$ particles, each one associated to a vertex in $\mathcal{V}$, a sequence of entangling operations between every pair of particles connected by an edge in $\mathcal{E}$ is applied to obtain the desired state \cite{graph_review}. 

\par Alternatively,  for DV systems, there exists a conceptually different approach: Every finite-dimensional graph state is known to be the unique ground state of a local, gapped {\it graph Hamiltonian}. By local we mean involving direct interactions among only a fixed number of particles; and by gapped we refer to $N$-independent finite energy difference between the ground and first-excited states. These Hamiltonians make adiabatic state creation possible: By engineering the interactions so as to effectively reproduce the Hamiltonian, the state can be prepared simply by first cooling down the system to zero temperature and then switching on the interactions so that the system is {\it adiabatically driven} toward the ground state of the graph Hamiltonian. The energy gap  imposes a threshold to the energy that the environment or erroneous manipulations have to pump into the system in order to drive it out of the ground state. This peculiarity provides the adiabatic  approach with an intrinsic robustness that the operational approach does not possess. In addition, the adiabatic  approach is naturally better suited for state preparations at large scales.  This explains the great theoretical effort devoted to finding local gapped Hamiltonians with universal resources for MBQC as ground states in the DV case,  where considerable progress has been achieved \cite{Hamiltonians}. On the other hand,  no such Hamiltonian has been reported for the CV domain. This is what this Letter presents.

\par We derive CV graph Hamiltonians $H_{G}$ that are quadratic, gapped, frustration-free (defined below), and of range two  (coupling only nearest or  next-to-nearest neighbors). The gap is independent on $N$ and is proportional to the squared inverse of a squeezing parameter $s$. The ground states are the  {\it CV Gaussian graph states} \cite{Schuch,Menicucci,Zhang,Ohliger}, which become universal resources for MBQC \cite{Menicucci}  at criticality $s\rightarrow\infty$. With Gaussian states as ground states, the fact that $H_{G}$ is quadratic---and therefore two-body---is no surprise. However, this is still in striking contrast to the case of qubits, where it is known that no state useful for MBQC can be the unique ground state of any two-body frustration-free Hamiltonian \cite{HamsNoGo}. 

\par With $H_{G}$ at hand, it is now also possible to assess the properties of  these bosonic systems in the realistic case of thermal equilibrium at nonzero temperatures. For example,  we show that the correlations  across any multipartition of the total thermal state are  determined exactly by  those of the boundary subsystem in a thermal state at the same temperature. By correlations, we refer to those measured by  {\it any} quantifier invariant under local unitary transformations, entanglement being arguably the most prominent example thereof. The boundary subsystem is composed by the bosons lying at the boundary of the multipartition, and is typically much smaller than the total system. So,  a considerable decrease in computational effort is gained for the correlations calculation. In turn, this automatically delivers a  correlation area law \cite{Eisert08}, even at criticality.  With this, in addition, we prove that for any $s<\infty$,   these systems typically display thermal bound entanglement, even in the thermodynamic limit $N\rightarrow\infty$. 
\par{\it CV graph Hamiltonians.---}Let us start by recalling the operational definition of graph states for CV quantum modes (qumodes) \cite{Menicucci,Zhang}, which is completely analogous to that of the DV case: First, for each $j\in\mathcal{V}$, initialize qumode $j$ in the zero-eigenvalue eigenstate $\ket{0}_{p_j}$  of the momentum-quadrature operator $\hat{p}_{j}$. The computational basis is taken as that of the eigenstates $\ket{v_j}_{q_j}$ of the position-quadrature operator $\hat{q}_{j}$, with $[\hat{q}_j,\hat{p}_k]=i\delta_{jk}, \forall\ j,k \in\mathcal{V}$ (we take $\hbar\equiv 1$ throughout). The two bases are related via  the Fourier transform $F_j$: $\ket{v_j}_{p_j}\equiv\frac{1}{\sqrt{2\pi}}\int_{\mathbb{R}}du_je^{iu_jv_j}\ket{u_j}_{q_j}=F_j\ket{v_j}_{q_j}$ . Our starting point  $\ket{0}_{p_j}$ is thus the uniform superposition of all computational states, exactly as in the DV case. Second, for each $\{j,k\}\in\mathcal{E}$, apply a maximally-entangling controlled-$Z$ gate  $CZ_{jk}\equiv e^{i\hat{q}_{j}\otimes\hat{q}_k}$ on  neighboring qumodes $j$ and $k$. The resulting state is the CV graph state  
\begin{equation}
\label{graph state}
\ket{G}=CZ\ket{0}_p,
\end{equation}
where $CZ$ (without subindices) is a short-hand notation for $\prod_{\{j,k\}\in\mathcal{E}} CZ_{jk}$ and $\ket{v}_p\equiv\bigotimes_{j\in\mathcal{V}}\ket{v_j}_{p_j}$, with $v$ (without subindex) standing for the multi-index $(v_1, \ ...\ v_N)\in\mathbb{R}^N$. In Eq. \eqref{graph state} (and throughout the Letter) we use  ``$G$" to represent ``$G_{(\mathcal{V},\mathcal{E})}$", unless explicitly indicated. Finally, it is also convenient to introduce  the  {\it nullifiers} 
\begin{equation}
\label{nullifiers} 
\hat{N}_i\equiv \hat{p}_i-\sum_{j\in\mathcal{N}_i}\hat{q}_j,
\end{equation}
with $\mathcal{N}_i$  all the neighbors of $i$, whose null-eigenvalue eigenstates are the CV graph states: $\hat{N}_i\ket{G}=0,\ \forall i\in\mathcal{V}$ \cite{Menicucci}. The latter is a necessary and sufficient condition to univocally specify state \eqref{graph state}. 

\par Momentum eigenstate $\ket{0}_{p_j}$ can in turn be obtained by infinitely squeezing the  vacuum coherent state $\ket{0}_{j}$:  $\ket{0}_{p_j}\equiv\lim_{s\rightarrow\infty}S_j(s)\ket{0}_{j}$. The action of the unitary squeezing operator  $S_j(s)\equiv e^{i\ln(s)(\hat{q}_{j}\hat{p}_{j}+\hat{p}_{j}\hat{q}_{j})/2}$  is to squeeze the position quadrature by a factor of $s$ and to stretch the conjugate momentum quadrature by a factor of $1/s$: $S^{\dagger}_j(s)\hat{q}_{j}S_j(s)\equiv \hat{q}_{j}s$ and $S^{\dagger}_j(s)\hat{p}_{j}S_j(s)\equiv \hat{p}_{j}/s$. Thus, states \eqref{graph state} can  be obtained from the vacuo $\ket{0}\equiv\bigotimes_{j\in\mathcal{V}}\ket{0}_{j}$ in the following way: $\ket{G}=\lim_{s\rightarrow\infty}CZS(s)\ket{0}$, being $S(s)\equiv\bigotimes_{i\in\mathcal{V}}S_i(s)$, with $s\equiv(s_1, \ ...\ s_N)$. In a more general way, finitely-squeezed Gaussian graph states are defined as 
\begin{equation}
\label{Gaussian graph state}
\ket{G_{s}}=U(s)\ket{0}\equiv CZS(s)\ket{0},
\end{equation} 
where  $U(s)\equiv CZS(s)$ has been introduced.
\par Now, consider the eigen-equation of $N$ free, non-interacting harmonic oscillators in the ground state, 
\begin{equation}
\label{Vacuum}
\hat{H}_0\ket{0}\equiv\sum_{i\in\mathcal{V}}\frac{\omega_i}{2}(\hat{q}_i^2+\hat{p}_i^2)\ket{0}=E_0\ket{0},
\end{equation}
with Hamiltonian $\hat{H}_0$, angular frequencies $\omega_i>0$, and ground-state energy $E_0\equiv\sum_{i\in\mathcal{V}}\omega_i/2$. Next apply the unitary operator $U(s)$ to both sides of  \eqref{Vacuum} from the left:
\begin{eqnarray}
\label{Vacuumsqueezed}
U(s)\sum_{i\in\mathcal{V}}\frac{\omega_i}{2}(\hat{q}_i^2+\hat{p}_i^2)U^{\dagger}(s) U(s)\ket{0}
\equiv E_0\ket{G_{s}}.
\end{eqnarray}
Invoking the  quadrature transformations under squeezing above we can rewrite the last equation as $CZ\sum_{i\in\mathcal{V}}\frac{\omega_i}{2}(\hat{q}_i^2/s^2+\hat{p}_i^2 s^2)CZ^{\dagger}\ \ket{G_{s}}=E_0\ket{G_{s}}$, where $S^{\dagger}(s)=S(1/s)$ has been used. The remaining controlled-$Z$ gates commute  with all position operators but transform each momentum operator as  $CZ^{\dagger}_{ij}\hat{p}_jCZ_{ij}=\hat{p}_i+\hat{q}_j$. Thus it is  $CZ\hat{p}_jCZ^{\dagger}=\hat{N}_j$. Equation \eqref{Vacuumsqueezed} then takes the form $\sum_{i\in\mathcal{V}}\frac{\omega_i}{2}(\hat{q}_i^2/s^2+\hat{N}_i^2 s^2)\ket{G_{s}}=E_0\ket{G_{s}}$, which  can be renormalized to the more convenient form
\begin{eqnarray}
\label{Vacuumsqueezedrenorm}
\hat{H}_G(s)\ket{G_{s}}=\frac{E_0}{s^2}\ket{G_{s}},\text{ with}\\
\label{Vacuumsqueezed2renorm}
\hat{H}_G(s)\equiv\sum_{i\in\mathcal{V}}\frac{\omega_i}{2}(\hat{q}_i^2/s^4+\hat{N}_i^2).
\end{eqnarray}
Equation \eqref{Vacuumsqueezedrenorm} constitutes a new ground-state eigen equation, with \eqref{Vacuumsqueezed2renorm} as the new Hamiltonian, \eqref{Gaussian graph state} as the new ground state, and $\frac{E_0}{s^2}$ as the new  ground-state energy. 

\par Hamiltonian \eqref{Vacuumsqueezed2renorm} is in turn the desired graph Hamiltonian. The  natural appearance of the nullifier operators in it is remarkable. As anticipated, the two-body character of $\hat{H}_G(s)$ is encapsulated in its quadratic form. In view of Eq. \eqref{nullifiers}, it is clear that direct couplings only take place between nearest, or next-to-nearest, neighbors. In addition, each and all of the terms in \eqref{Vacuumsqueezed2renorm} commute, which implies that the Hamiltonian is frustration-free (meaning that the ground state minimizes the energy of each local term in the sum). Furthermore, the gap between the ground and first-excited states is that of $\hat{H}_0$ (${\omega_i}_{min}\doteq\min_{\{i\in\mathcal{V}\}}\omega_{i}$) consistently renormalized: ${\omega_i}_{min}/s^2$. At infinite squeezing the gap vanishes and the system is then called critical. Interestingly enough however, its thermal states always satisfy a correlation area law, as we show below. In addition, at criticality the graph Hamiltonian \eqref{Vacuumsqueezed2renorm} acquires an even simpler form: $\hat{H}_G\equiv\lim_{s\rightarrow\infty}\hat{H}_G(s)=\sum_{i\in\mathcal{V}}\frac{\omega_i}{2}\hat{N}_i^2$.

\par The derivation of  \eqref{Vacuumsqueezed2renorm}  not only delivers the desired Hamiltonian but also comes with the interesting byproduct of readily giving the symplectic transformation that takes the Hamiltonian to its {\it normal mode decomposition}. In this case, $U(s)$ is the unitary representation of such transformation and maps our Hamiltonian to that of a collection of noninteracting harmonic oscillators:
\begin{eqnarray}
\label{normalmode}
U^{\dagger}(s)\hat{H}_G(s)U(s)\equiv S^{\dagger}(s)CZ^{\dagger}\hat{H}_G(s)CZS(s)=\hat{H}_0/s^2,
\end{eqnarray}
with $\hat{H}_0$ given in \eqref{Vacuum}. What is more, the same unitary transformation delivers also the {\it Gaussian nullifiers}. That is, $N$ commuting operators with the Gaussian graph state \eqref{Gaussian graph state} as their (unique) mutual eigenstate of null eigenvalue. To see this, instead of  eigen equation \eqref{Vacuum}, start by $\hat{a}_j\ket{0}\equiv0$, with $\hat{a}_j\equiv(\hat{q}_j+i\hat{p}_j)/\sqrt{2}$ the annihilation operator of the $j$th qumode, and  with the same reasoning as above arrive at the Gaussian nullifiers $\hat{N}_j(s)\equiv -i\hat{q}_j/s^2+\hat{N}_j$, where $\hat{N}_j$ is the $j$th  nullifier \eqref{nullifiers}.
\par {\it Thermal Gaussian graph states.---}Once we have obtained Hamiltonian \eqref{Vacuumsqueezed2renorm} we can now  consider thermal  Gaussian graph states, which are defined in the usual way:
\begin{equation}
\label{rho thermal}
\rho_{G,T}\equiv\frac{e^{-\hat{H}_{G}/ T}}{\tr\big[e^{-\hat{H}_{G}/ T}\big]},
\end{equation}
where  $T$ is the temperature of the system's thermal bath (Boltzmann's constant is set as unit).

\par We are now in a position to study the correlations $C(\rho_{G,T})$ across any multipartition of thermal state \eqref{rho thermal}, with $C$ any arbitrary correlation quantifier  invariant under local unitary transformations. For its evaluation we first decompose  the symplectic unitary as 
$U(s)\equiv CZ S(s)=CZ_{\mathcal{X}}S_{\mathcal{Y}}(s)\otimes CZ_{\overline{\mathcal{X}}}S_{\overline{\mathcal{Y}}}(s)\equiv U_{\mathcal{Y}}(s)\otimes U_{\overline{\mathcal{Y}}}(s)$. 
$\mathcal{X}\subseteq\mathcal{E}$ is the set of {\it boundary-crossing edges}  \cite{dan1}, those that connect vertices belonging to different subpartitions. The latter vertices in turn compose the set  $\mathcal{Y}\subseteq\mathcal{V}$ of {\it boundary
vertices}  \cite{dan1}. The two sets together constitute the {\it boundary subgraph} $G_{(\mathcal{Y},\mathcal{X})}$, and the rest   $G_{(\overline{\mathcal{Y}},\overline{\mathcal{X}})}$, with $\overline{\mathcal{Y}}\equiv\mathcal{V}/\mathcal{Y}$ and $\overline{\mathcal{X}}\equiv\mathcal{E}/\mathcal{X}$, is called the {\it nonboundary subgraph}. With this, we notice that $\rho_{G,T}\equiv U_{\overline{\mathcal{Y}}}(s)\rho_{G_{(\mathcal{Y},\mathcal{X})},T}\otimes\rho_{0_{(\overline{\mathcal{Y}})},T}U^{\dagger}_{\overline{\mathcal{Y}}}(s)$. In the last,  
$\rho_{G_{(\mathcal{Y},\mathcal{X})},T}$ is a thermal state of the boundary subsystem, defined as in Eq. \eqref{rho thermal} but with respect to the boundary subgraph Hamiltonian $H_{G_{(\mathcal{Y},\mathcal{X})}}\equiv\sum_{i\in\mathcal{Y}}\frac{\omega_i}{2}(\hat{q}_i^2/s^4+{\hat{N}_{\mathcal{Y}i}}^2)$. Here, $\hat{N}_{\mathcal{Y}i}\equiv \hat{p}_i-\sum_{j\in\mathcal{N}_{\mathcal{Y}i}}\hat{q}_j$ is the $i$th nullifier corresponding to the boundary subgraph -- the same as in Eq. \eqref{nullifiers}  but with the sum running over the set  $\mathcal{N}_{\mathcal{Y}i}\equiv\mathcal{N}_i\cap\mathcal{Y}$  of the neighbors of $i$ in $\mathcal{Y}$ --. In turn, $\rho_{0_{(\overline{\mathcal{Y}})},T}$ is a thermal state of the nonboundary subsystem with respect to the decoupled harmonic Hamiltonian $H_{0_{(\overline{\mathcal{Y}})}}\equiv\sum_{i\in\overline{\mathcal{Y}}}\frac{\omega_i}{2}(\hat{q}_i^2+\hat{p}_i^2)$. Notice that  $H_{G_{(\mathcal{Y},\mathcal{X})}}$ and $H_{0_{(\overline{\mathcal{Y}})}}$ commute. 

\par Now, by definition, $U_{\overline{\mathcal{Y}}}(s)$ is a local unitary operation with respect to the considered multipartition, so we can disregard it as for what the correlation evaluation concerns. Once $U_{\overline{\mathcal{Y}}}(s)$ is omitted,  the boundary and nonboundary subsystems are left in the product $\rho_{G_{(\mathcal{Y},\mathcal{X})},T}\otimes\rho_{0_{(\overline{\mathcal{Y}})},T}$, so all the correlations across the multipartition are concentrated in its boundary: 
\begin{equation}
\label{thermalent}
C(\rho_{G,T})=C(\rho_{G_{(\mathcal{Y},\mathcal{X})},T}).
\end{equation}

\par Thermal states of one-dimensional bosonic chains governed by some specific families of finite-ranged quadratic Hamiltonians are known to satisfy an entanglement area law for some particular entanglement quantifiers  \cite{Eisert08}. An area law is said to be satisfied when  the correlations across a multiparition scale at most with the size of its boundary. Equivalence \eqref{thermalent} gives us the fully general statement for thermal Gaussian graph states: they obey an {\it area law for any geometry and any local-unitary invariant correlation}. Not only that, it gives us much more refined information  for it reduces the correlation-evaluation problem of arbitrarily sized specimens  (for example macroscopic ones)   to one on the boundary subsystem, regardless of how well an area law is satisfied.
\begin{figure}[t]
\begin{center}
\includegraphics[width=1\linewidth]{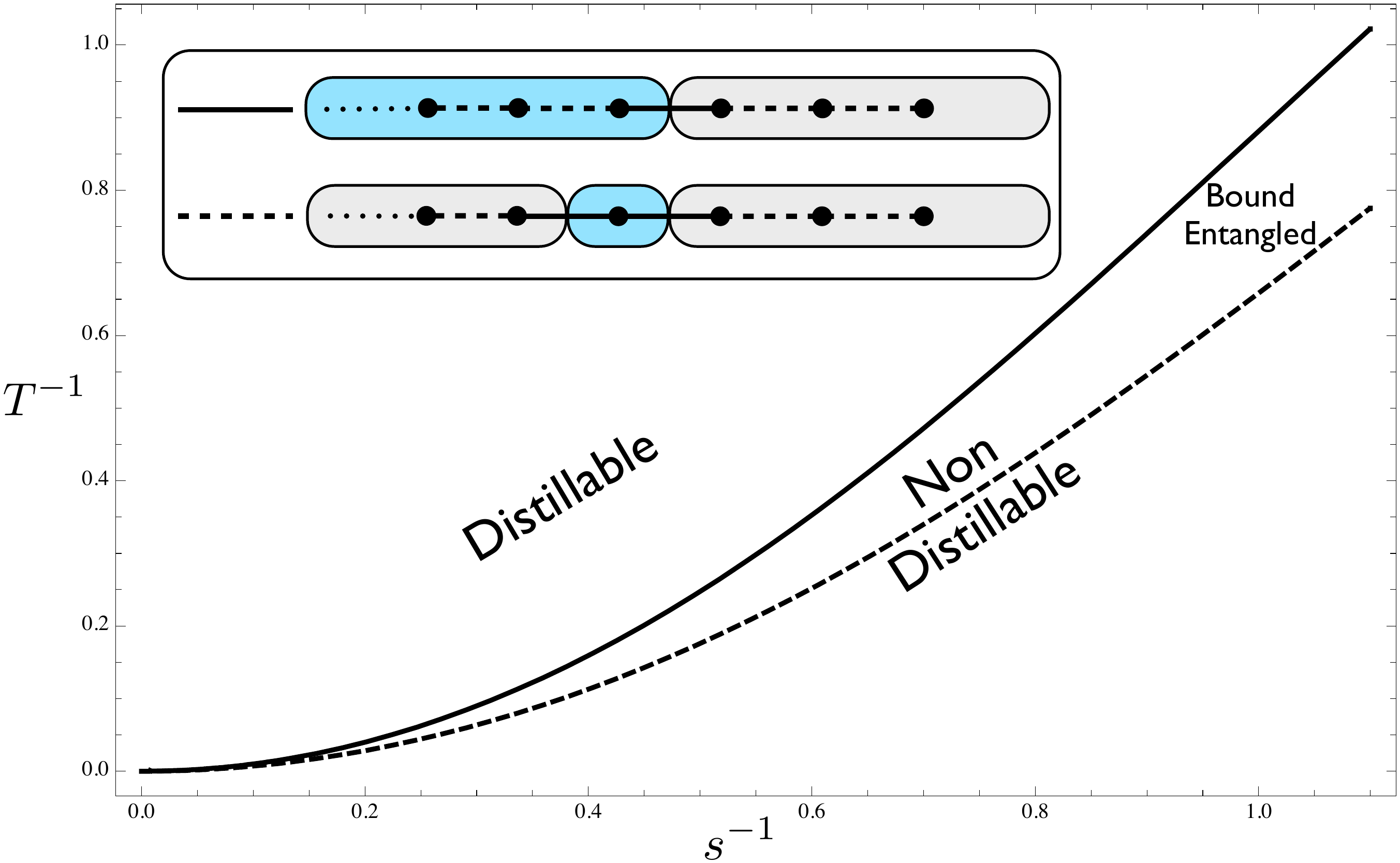}
\caption{High-temperature, high-squeezing region of the distillability phase diagram of thermal Gaussian graph states associated to arbitrarily large linear clusters. The solid curve represents the inverse critical temperature ${T_2}_{c}^{-1}=2\ \text{arccoth}\big(\sqrt{1+s^4}+s^2\big)$ (in arbitrary units) at which the negativity between any two contiguous blocks vanishes as a function of the inverse squeezing, and the dashed one the analogous ${T_3}_{c}^{-1}=2\ \text{arccoth}\big[\big(\sqrt{1+4s^4-2\sqrt{2}\sqrt{2s^8+s^4}}\big)^{-1}\big]$ for the bipartitions of any qumode versus the rest.  Between the two curves every qumode is entangled with all the rest but no entanglement can be locally extracted. \label{Diagram}}
\end{center}
\end{figure}
\par   In Fig. \ref{Diagram} for instance we have plotted the inverse critical temperatures at which the logarithmic negativities \cite{Vidal} of state \eqref{rho thermal} vanish as a function of the inverse squeezing, for the exemplary case of one-dimensional graphs and constant Hamiltonian couplings $\omega_i=\omega$. The solid curve corresponds to any block-to-block bipartition, and the dashed one to any one-mode-versus-the-rest bipartition [see inset of Fig. \ref{Diagram}[. By equivalence \eqref{thermalent}, the former is given simply by the critical temperature ${T_2}_c$ of a two-qumode thermal cluster state, and the latter by that of the bipartition of one qumode versus the other two in a three-qumode thermal linear-cluster state,  ${T_3}_c$. Both temperatures can be calculated analytically. For temperatures higher than ${T_2}_c$ not even two-mode entanglement can be (locally) distilled between any two qumodes in the chain, for if this were possible then there would be two contiguous blocks with positive negativity \cite{Ferraro,CavalcantiThermal}. On the other hand, for temperatures lower than (at least) ${T_3}_c$ every boson in the chain is entangled with all the rest. For ${T_2}_c\leq T<{T_3}_c$ thus, thermalization naturally drives  Gaussian 1 D graph states of any size to bound entanglement. This extends of course to higher-dimensional clusters and nonequal couplings, as in the DV case \cite{CavalcantiThermal}.

\par {\it Phase-space diffusions on CV graph states.---}In the limit of infinite squeezing, thermal states \eqref{rho thermal} become equivalent to the result of independent Gaussian diffusion along the $\hat{q}$ direction on pure states \eqref{graph state}. This interesting connection between (collective) thermalization and (independent) dephasing is the CV version of the one observed in qubit graph states \cite{CavalcantiThermal,Kay}. Additionally, also for $s\rightarrow\infty$, the evolution of correlations under noise processes described by arbitrary phase-space-shift maps can be monitored in terms of the boundary subsystem by translating the whole machinery for the study of  DV graph-state entanglement under Pauli maps developed in Ref. \cite{dan1}. This characterization is relevant  to the development of CV quantum error correction schemes, but it will be touched upon elsewhere.
\par {\it Discussions.---}As known, states \eqref{graph state} are as nonphysical idealizations as the free-particle states $\ket{0}_p$. However, universality for MBQC has only been proven for infinitely squeezed states \cite{Menicucci}. What is more, recent results \cite{Ohliger} show that large-scale MBQC with ``imperfect" physical states \eqref{Gaussian graph state} faces fundamental limitations that may only be circumvented with the full machinery of quantum error correction and fault tolerance \cite{Raussendorf07}. The latter is yet to be developed for CV systems, but  will presumably demand a very large overhead in resources. This in turn highlights the importance of local, short-ranged, gapped CV graph Hamiltonians, for it is precisely the large-scale scenario where the adiabatic approach  is specially well suited. On the other hand, states \eqref{Gaussian graph state} do yield universal resources for small-sized computations.

From a more applied viewpoint, our findings open new realistic venues for the physical realization of CV quantum computing beyond the standard optical approaches. In fact, the basic constituents of Hamiltonian \eqref{Vacuumsqueezed2renorm} ---namely, the couplings  $\hat{q}_i\otimes\hat{q}_j$ and $\hat{p}_i\otimes\hat{q}_j$---have already been demonstrated in technologically mature experimental platforms, as Coulomb crystals \cite{ions} and opto-mechanical resonators \cite{optomechanics}. In addition, this type of couplings have also been envisioned in coupled-microcavities arrays or superconducting waveguides \cite{Hartmann}. All these constitute examples of versatile and promising architectures where controlled geometrical arrangements of the desired interactions seem feasible in a very near future. 
\par {\it Note added.---}After conclusion of this work, another paper addressing CV graph Hamiltonians appeared \cite{Menicucci2}.
\par {\it Acknowledgments.---}L. A.  thanks D. Browne for reviving his interest in CV systems. This work was supported by  the European COMPAS project and the ERC Starting grant PERCENT, the Spanish MEC FIS2007-60182, Consolider-Ingenio QOIT and Juan de la Cierva projects, the Generalitat de Catalunya and Caixa Manresa.

\end{document}